\documentclass[a4paper,preprint,aps]{revtex4}

\usepackage{graphicx}
\usepackage{dcolumn}
\usepackage{bm}

\newcommand{\eqa}{\begin{equation}}
\newcommand{\eqz}{\end{equation}}
\newcommand{\eqma}{\begin{eqnarray}}
\newcommand{\eqmz}{\end{eqnarray}}

\begin{document}
\title{Anharmonic Force Fields and Thermodynamic Functions
using Density Functional Theory\footnote{Dedicated to Prof. Nicholas C. Handy FRS on the occasion of his 63rd birthday.}}
\author{A. Daniel Boese}
\affiliation{Institute of Nanotechnology, Forschungszentrum
Karlsruhe, P.O. Box 3640, D-76021 Karlsruhe, Germany}
\author{Wim Klopper}
\affiliation{Lehrstuhl f\"ur Theoretische Chemie, Institut f\"ur
Physikalische Chemie, Universit\"at Karlsruhe (TH), D-76128
Karlsruhe, Germany}
\author{Jan M. L. Martin}
\email{comartin@wicc.weizmann.ac.il} \affiliation{Department of
Organic Chemistry, Weizmann Institute of Science, IL-76100
Re\d{h}ovot, Israel}
\date{Received July 22, 2004; Revised November 4, 2004; Manuscript \# {\bf MP84509}}
\smallskip
\begin{abstract}
\indent The very good performance of modern density functional
theory for molecular geometries and harmonic vibrational frequencies
has been well established. We investigate the performance of density
functional theory (DFT) for quartic force fields, vibrational
anharmonicity and rotation-vibration coupling constants, and
thermodynamic functions beyond the RRHO (rigid rotor-harmonic
oscillator) approximation of a number of small polyatomic molecules.
Convergence in terms of basis set, integration grid and the
numerical step size for determining the quartic force field by using
central differences of analytical second derivatives has been
investigated, as well as the performance of various
exchange-correlation functionals. DFT is found to offer a
cost-effective approach with manageable scalability for obtaining
anharmonic molecular properties, and particularly as a source for
anharmonic zero-point and thermal corrections for use in conjunction
with benchmark {\it ab initio} thermochemistry methods.
\end{abstract}
\maketitle
\section{Introduction}

The Kohn-Sham formulation of Density Functional Theory (DFT) is
nowadays a commonly used tool in computational chemistry, being
frequently applied to thermodynamics, structures, harmonic
frequencies of various chemical compounds. Despite its successes in
the accurate prediction of most ground-state and some excited state
properties of molecules, its usefulness for calculating anharmonic
force fields is less firmly established.

Anharmonic force fields allow direct comparison between computed and
observed  spectroscopic transitions, as opposed to comparing
computed harmonic apples with observed anharmonic oranges, all the
while making `hand-waving' approximations about the importance of
anharmonicity. The cost of high-quality {\it ab initio} anharmonic
force field calculations rapidly becomes prohibitive as the number
of atoms increases, and the relatively low cost of DFT calculations
--- as well as the fairly routine availability of analytical second
derivatives --- makes them attractive potential alternatives.

Until about two years ago, DFT anharmonic force field studies were
rather scarce.
Dressler and Thiel studied H$_2$O, F$_2$O, and
CH$_3$F using several generalized gradient approximation (GGA)
functionals\cite{thiel}. Detailed studies had been done for
diatomics\cite{Hassanzedeh,Sinnokrot}, with the respective authors
claiming very high accuracy. Concerning larger molecules, the full
quartic force field of ammonia\cite{anharmhandy1},
benzene\cite{anharmhandy2} and diazomethane\cite{Baraille} have been
studied with a hybrid density functional.

In the last couple of years, an increasing number of anharmonic
force fields using density have been published and several groups
have been working on this subject. Noteworthy are the activities in
the Handy group with studies of furan, pyrrole and
thiophene\cite{Rudolf} and phosphorus pentafluoride\cite{Tew}, which
in fact yielded even more accurate results on these medium-sized
organic systems than suggested by the small-molecule validation
studies cited above. In addition to this, the group of
Hess\cite{Neugebauer} (using the B3LYP\cite{B3LYP} and
BP86\cite{BP86} functionals) and Barone\cite{Baroneval}
(using B3LYP) have published validation studies on a small number of
mainly triatomic molecules. Furthermore, two of us published a
detailed study on the azabenzene series, also exploring the
possibility of combining DFT anharmonic force fields with coupled
cluster geometries and harmonic frequencies\cite{Azabenzenes}.
Barone carried out a similar study on the azabenzenes using a
different functional and basis set\cite{BaroneAzabenzenes}.

Another potential application of interest concerns high-accuracy
computational thermochemistry. Computed molecular atomization
energies require zero-point vibrational energies (ZPVEs) and thermal
corrections. As molecules grow, ZPVEs make up an increasingly
important part of the molecular binding energy --- e.g., 62.08
kcal/mol in benzene\cite{benzeneW2} --- and approximations about the
anharmonic contribution to the ZPVE introduce ever larger sources of
potential error. This issue becomes especially acute with
nonempirical extrapolation-based methods like W1, W2, and W3
theory\cite{W2,W3} or explicitly correlated methods like
CC-R12\cite{KlopperReview}, which can fairly routinely yield
atomization energies to sub-kcal/mol accuracy. Thermal corrections
at room temperature can reasonably be expected to be reproduced
fairly well for semirigid molecules, but in applications where
high-temperature data are important (e.g., combustion modeling),
vibrational anharmonicity, rovibrational coupling, and centrifugal
distortion  significantly affect thermodynamic functions (see, e.g.,
Refs.\cite{h2o,cf2}). All the required molecular properties can be
readily obtained by vibrational perturbation theory when calculating
the quartic force field of the respective
compound\cite{rovib1,rovib2,rovib3,rovib4,rovib5}.

Since the computation of a quartic force field using density
functional theory is unlikely to require more computer time than
W$n$ or explicitly correlated methods, it could become a routine
step in such studies. It was previously shown\cite{W2book} that the
use of accurate ZPVEs from large-scale ab initio anharmonic force
field calculations, instead of scaled B3LYP/cc-pVTZ harmonic
frequencies, improved the mean absolute error of W2 theory over the
W2-1 set of 28 small molecules (with very precisely known
experimental binding energies) from 0.30 to 0.23 kcal/mol, and the
maximum absolute error from 0.78 to 0.64 kcal/mol. This can be
compared to the error when including up to connected quadruple
excitations in the extrapolation scheme, where the mean absolute
error (over a sample additionally including some molecules beset
with nondynamical correlation) is reduced from 0.40 (W2) to 0.22
(W3) kcal/mol.

Assuming that we would like to achieve the best accuracy possible,
we ought to be able to mitigate the error introduced by the ZPVE by
using quartic DFT force fields to calculate the anharmonic
corrections.

Of course, other approaches than vibrational perturbation theory
could be used in conjunction with density functional methods, such
as the VSCF-CC
method\cite{Gerber1,Gerber2,Gerber3,Gerber4,Hirao1,Christiansen}.
However, as these methods are presently $J=0$ (i.e., rotational
ground state) methods, the effects of rovibrational coupling and
centrifugal distortion on the thermodynamic functions would have to
be neglected.

In the present work, we report on a detailed validation study of DFT
methods for all these properties, and will consider the dependence
of their accuracy on the exchange-correlation functional, the basis
set, the quality of the numerical integration grid, and the step
size used in numerical differentiation.

\section{Computational Details}

Following the approach first proposed by Schneider and
Thiel\cite{rovib4}, a full cubic and a semidiagonal quartic force
field are obtained by central numerical differentiation (in
rectilinear normal coordinates about the equilibrium geometry) of
analytical second derivatives. The latter were obtained by means of
locally modified versions of {\sc gaussian} 98\cite{g98}; modified
routines from {\sc cadpac}\cite{Cadpac} were used as the driver for the
numerical differentiation. routine. In this approach, the potential
energy surface is expanded through quartic terms at the global
minimum geometry like:
\begin{eqnarray}
V=\frac{1}{2}\sum_{i}\omega_i q^2_i +
\frac{1}{6}\sum_{ijk}\phi_{ijk}q_i q_j q_k +
\frac{1}{24}\sum_{ijk}\phi_{ijkk}q_i q_j q_k q_l
\end{eqnarray}
where the $q_i$ are dimensionless rectangular normal coordinates,
$\omega_i$ are harmonic frequencies, and $\phi_{ijk}$ and
$\phi_{ijkl}$ third and fourth derivatives with respect to the $q_i$
at the equilibrium geometry.

All the force fields have been analyzed by means of the
{\sc spectro}\cite{Spectro} and {\sc polyad}\cite{Polyad} rovibrational
perturbation theory programs developed by the Handy and Martin
groups, respectively.

In all cases, when strong Fermi resonances lead to band origins
perturbed more than about 2 cm$^{-1}$ from their second-order
position, the deperturbed values are reported and resonance matrices
diagonalized to obtain the true band origins. Rotational constants
were similarly deperturbed for strong Coriolis resonances.

Thermodynamic functions beyond the harmonic approximation are
obtained by means of the integration of asymptotic series method as
implemented in the NASA PAC99 program of McBride and
Gordon\cite{PAC99}.

\section{Results and discussion}

To avoid issues with experimental data, such as problems with the
assignment of the spectra as has been reported in our previous
study\cite{Azabenzenes}, we decided to solely compare our results to
{\it ab initio} (coupled cluster) data of quartic force fields which
have been published for molecules with more than two atoms. Here, 17
molecules, namely C$_2$H$_2$\cite{c2h2}, C$_2$H$_4$\cite{c2h4},
CCl$_2$, CF$_2$\cite{cf2}, CH$_2$NH\cite{ch2nh}, CH$_2$\cite{ch2},
CH$_4$\cite{ch4}, H$_2$CO\cite{h2co}, H$_2$O\cite{h2o},
H$_2$S\cite{h2s}, HCN\cite{hcn}, N$_2$O\cite{n2o}, NH$_2$\cite{nh2},
PH$_3$\cite{ph3}, SiF$_4$\cite{sif4}, SiH$_4$\cite{sih4}, and
SO$_2$\cite{so2}, have been included with some of the older data
being recalculated with a more extended basis set.

\subsection{Numerical quadrature}

First of all, we have to investigate the dependence on the
integration grid. We shall restrict ourselves to the HCTH/407
functional\cite{HCTH407}, using a TZ2P basis set for the H$_2$O,
SO$_2$ and N$_2$O molecules together with a step size of 0.02 a.u.
for the central differences.
The grids considered here are direct products of Euler-Maclaurin
radial grids\cite{Han93}
ranging from 75 to 400 points,
with Lebedev angular grids\cite{Lebedev} ranging from 194 to 974
points. In addition, we considered some `pruned' grids, in which the
radial coordinate is divided into five zones (inner core, inner
valence, valence, outer valence, and long-range) and different
angular grid densities are used for each zone. The `SG1', 'Medium',
'Fine', and 'Ultrafine' standard grids in {\sc gaussian} 9x correspond to
pruned (50,194), (75,194), (75,302), and (99,590) grids,
respectively. Inter-atom partitioning was done according to
Ref.\cite{SSF}.

\begin{figure}
\begin{tabular} {c}
\includegraphics[width=7cm,angle=270]{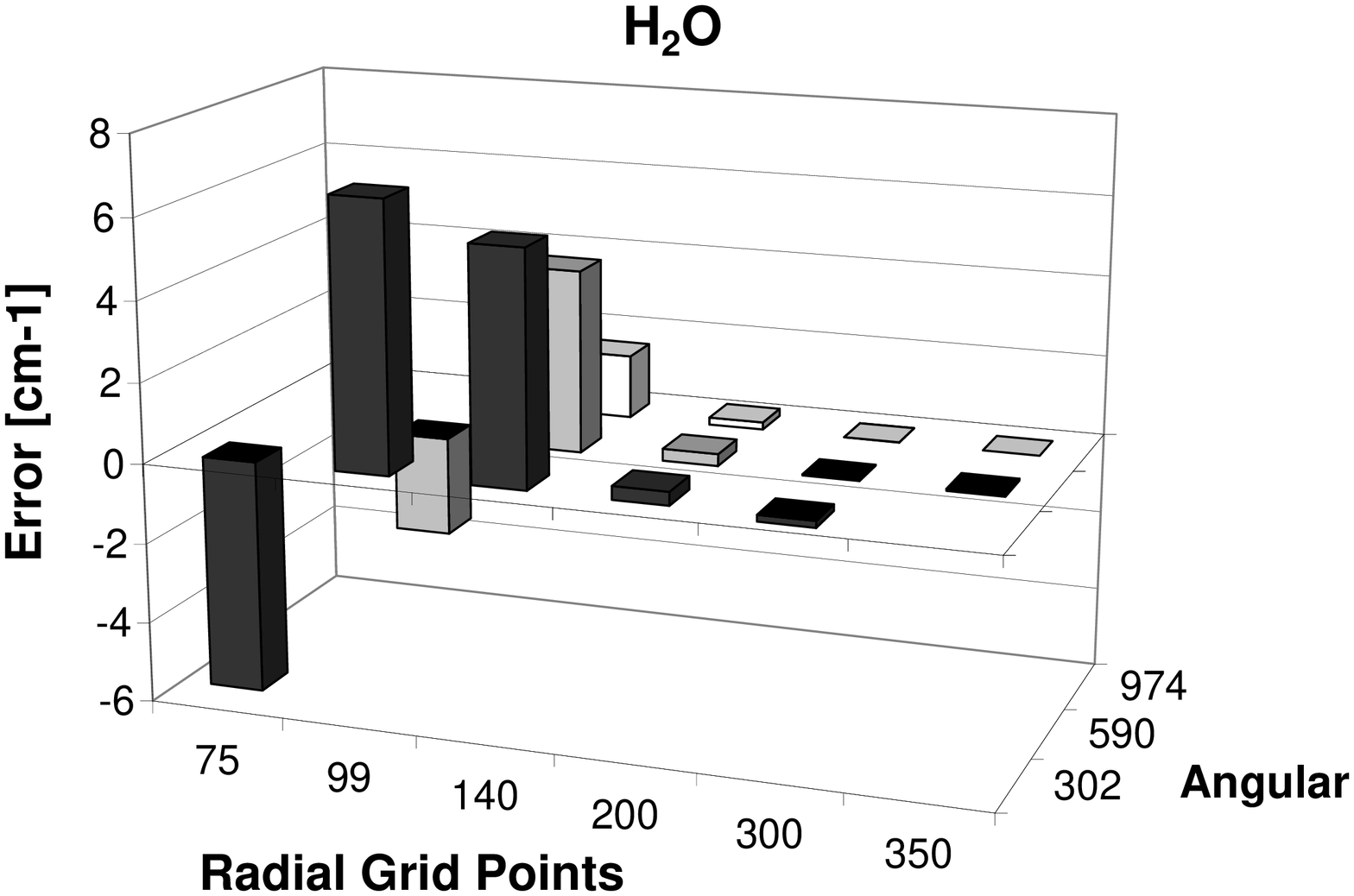} \\
\\
\includegraphics[width=7cm,angle=270]{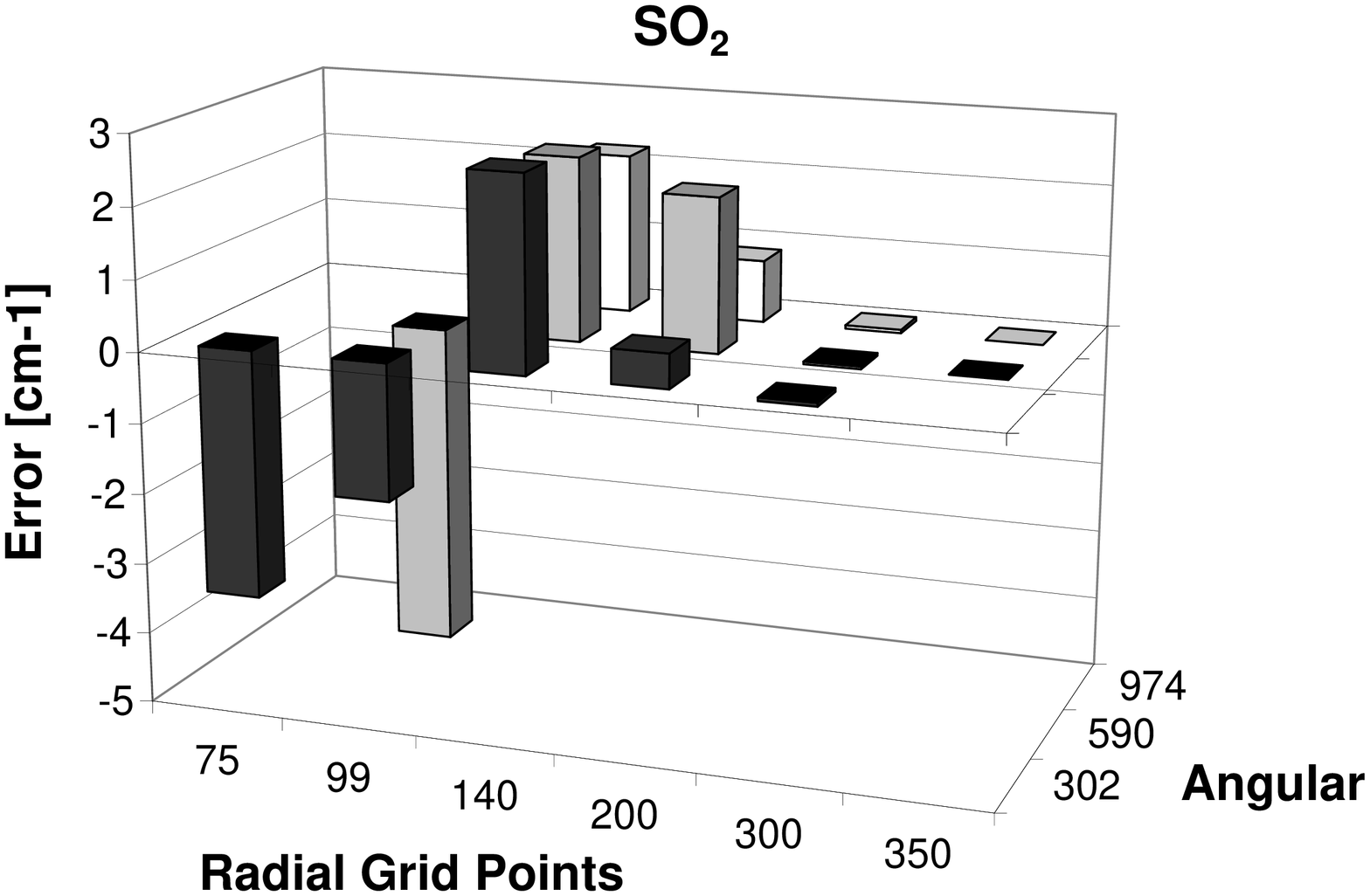} \\
\\
\includegraphics[width=7cm,angle=270]{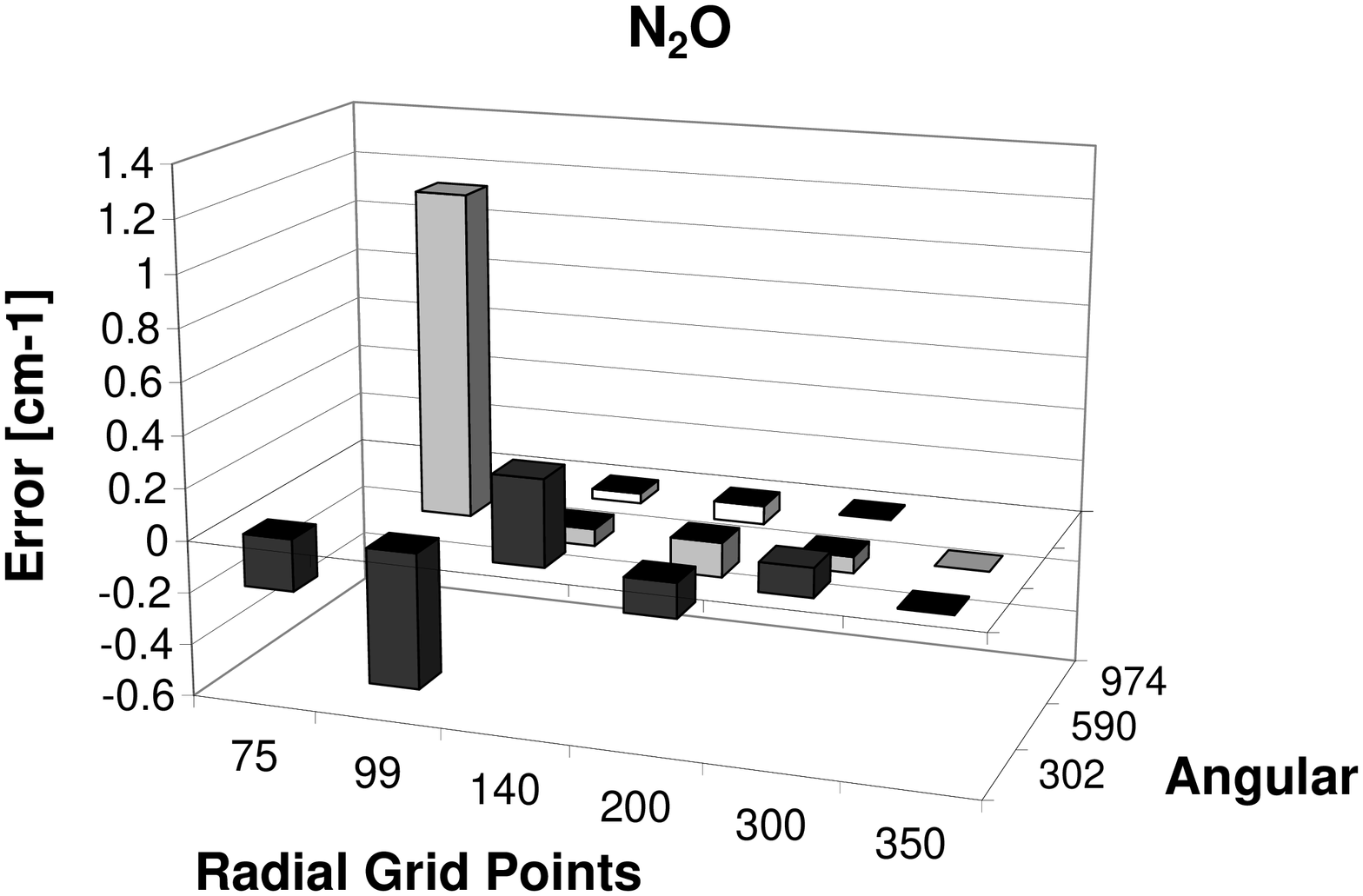} \\
\end{tabular}
\caption{\label{fig1}Boese et al}
\end{figure}

Let us first consider the error in the off-diagonal stretching
anharmonicity, relative to our largest grid (unpruned
400$\times$974). Unfortunately, sizable errors are seen for the
standard {\sc gaussian}-type grids, both due to their small intrinsic size
and to the pruning. Hence, they cannot be recommended for this type
of calculation, since these values are quite significant for the
calculation of the thermodynamic functions and fundamental
frequencies. The results are shown in Figure \ref{fig1}a to Figure
\ref{fig1}c. As it is apparent from these figures, a very large
number of radial grid points is needed for the accurate description
of the quartic force field. From the values in Figure 1, we conclude
that a 200$\times$974 grid for our calculations if the molecules
include only first-row atoms, and the 300$\times$974 grid otherwise
will be sufficient. We modified the code to add two additional
pruned grids, 140$\times$974 and 199$\times$974 (which are invoked
by new keywords 'Grid=Huge' and 'Grid=Insane', respectively). For
these intrinsically finer-meshed grids, pruning deteriorates
numerical precision less, and if 1 cm$^{-1}$ precision is
sufficient, the `Huge' grid will be generally adequate for
first-row, and  the `Insane' grid for second-row molecules.

Unfortunately, Gaussian does permit using different radial grids on
different atoms, although such variation can be somewhat simulated
by manipulation of the pruning zone boundaries. It is however
possible to use a different (coarser) grid in the CPKS
(coupled-perturbed Kohn-Sham) steps, and here we were able to reduce
grid size to as small as SG1 (i.e., pruned 75$\times$194), thus
reducing overall computational cost. For instance, for methylene
imine CH$_2$NH, this approximation reduced CPU time by half, while
the computed fundamental frequencies change by less than 0.2
cm$^{-1}$. The importance of using a very large {\em
energy+gradients} grid is apparent for this molecule as well: Even a
relatively large 99$\times$590 (or in its pruned version, `Ultrafine')
grid will yield errors for some frequencies of more than 8
cm$^{-1}$, and the 140$\times$590 grid errors of 1 cm$^{-1}$. We
expect similar behavior for other quantum chemical program systems,
where grid sizes will likewise need to be increased well beyond what
is normally required. Needless to say, errors are further
exacerbated if one of the fundamentals affected by grid error is
involved in a strong Fermi resonance.
For example, in CH$_2$NH, the change in going from a pruned
75$\times$302 to an unpruned 140$\times$590 CPHF grid changes
the eigenvalues of the fundamentals affected by Fermi resonances by about
ten cm$^{-1}$. The deperturbed values, however, change by less than 0.5 cm$^{-1}$.
Although this is clearly undesirable (and an inherent
weakness of the method), in such a situation it becomes unclear which
grid is actually the best to use, since even at large grid sizes the changes
are significant.

\subsection{Basis set}

\begin{figure}
\includegraphics[width=12cm,angle=270]{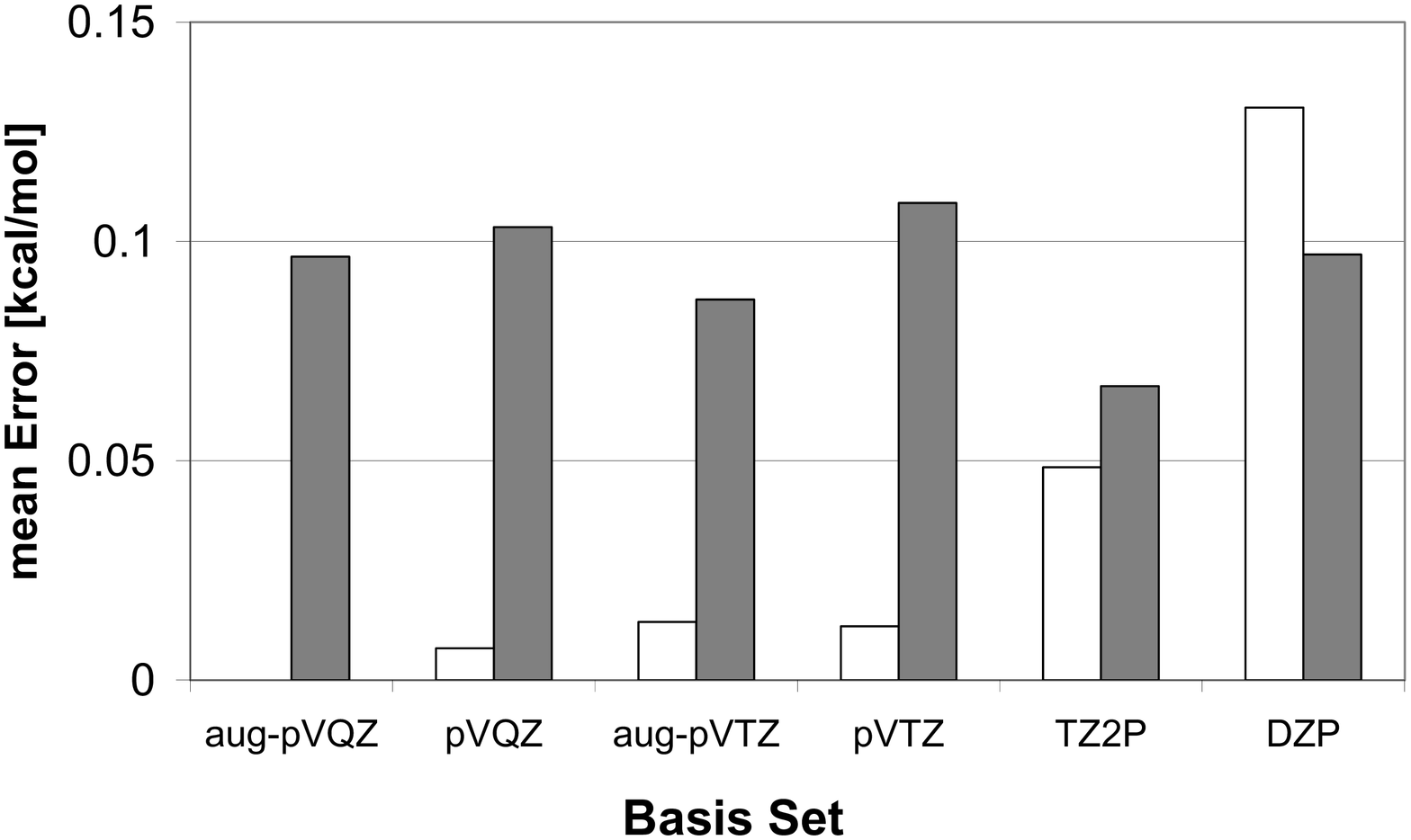} \\
\caption{\label{fig2}Boese et al}
\end{figure}

For the basis sets investigated, we have used Dunning's cc-pV$n$Z
and aug-cc-pV$n$Z correlation consistent basis sets for the first
row\cite{PVXZfirst}, and the cc-pV($n$+d)Z and aug-cc-pV($n$+d)Z
basis sets of Wilson, Peterson, and Dunning for the second row. (The
latter include additional high-exponent $d$ functions, which have
been shown to be important\cite{so2} for spectroscopic constants of
molecules in which a second-row atom is surrounded by one of more
highly electronegative first-row atoms.) As for smaller basis sets,
we considered the TZ2P\cite{TZ2P} and DZP \cite{DZP} basis sets.
(Note that the TZ2P version used by most groups actually includes a
third $d$ function for second-row atoms.) In all, we investigated
the aug-cc-pVQZ, cc-pVQZ, aug-cc-pVTZ, cc-pVTZ, TZ2P and DZP basis
sets, using the B97-1 hybrid functional\cite{HCTH93} and a step size
of 0.02 Bohr. Here, we compare to the largest basis set
(aug-cc-pVQZ), which for DFT calculations of these properties is
close to the Kohn-Sham basis set limit. In figure \ref{fig2}, now
the mean errors of the zero-point energies are shown for the
C$_2$H$_4$, H$_2$O, SO$_2$ and N$_2$O molecules, in addition to the
overall error of the functional compared to coupled cluster data.
The latter data shows when the basis error becomes significant
enough to affect the error made by the functional itself. This
happens only to the DZP basis set,
although its overall mean error compared to the reference data is
not larger than for the largest basis set employed. For the TZ2P
basis, there seems to be some error compensation for this
functional. Basically, the error made by the cc-pVTZ basis set is
still very small compared to the aug-cc-pVQZ basis set, with the
TZ2P basis set error still being significantly smaller than the
functional error and the DZP basis set error being comparable. Thus,
from these results, either the TZ2P or cc-pVTZ basis sets seem
preferable, with the DZP basis set being an option for the very
largest molecules.

\subsection{Step size for numerical differentiation}

As a third variable, an optimal numerical step size has to be
determined. In a numerical derivatives calculation, it always
represents a compromise between discretization error and roundoff
error. All calculations have been done using an unpruned
200$\times$590 grid for CH$_2$NH, a molecule which proved to be very
much affected by step size issues.
For the first three modes, we compare to the deperturbed values to
circumvent the difficulties mentioned above.
In order to reduce roundoff error
as much as possible, the KS and CPKS equations were basically
converged to machine precision.

We considered five different approaches to the step size.
All of them have been done along the unnormalized Cartesian 
displacement vectors of the mass weighted normal coordinates, 
such as\cite{Gaussianpages}:
\begin{eqnarray}
\vec{r}=\vec{r}+q_{step}(i)\times\vec{l}
\end{eqnarray}
All stepsizes $q_{step}(i)$ are in $bohr\times amu$.
In the first, a constant `one size fits all' displacement was made in all
modes:
\begin{eqnarray}
q_{step}(i)=\hbox{constant}
\end{eqnarray}

In the second, we employed a variable step size proportional to
the square root of the reduced mass for each particular vibration:
\begin{eqnarray}
q_{step}(i)=\hbox{constant}\times\sqrt{\frac{\mu}{\hbox{amu}}}\label{eq:method2}
\end{eqnarray}
This corresponds to normalising the displacement vector to have the
same stepsize for different isotopes.

The third approach is to choose a step size such that within the
harmonic oscillator approximation, the displacement in that
particular mode causes a certain given energy change, e.g., 1
millihartree as in Ref.\cite{anharmhandy2}

The fourth approach is to choose an unnormalised stepsize dependent
on the force constant $k$. The is achieved by dividing through the
reduced mass and the frequency of the corresponding mode:
\begin{eqnarray}
q_{step}(i)=\hbox{constant}\times\frac{1}{\sqrt{\frac{\mu}{\hbox{amu}}}}\times\frac{1000~\hbox{cm}^{-1}}{\omega(i)}
\propto \hbox{constant}\times\frac{1}{\sqrt{k}}\label{eq:method4}
\end{eqnarray}

Finally, in the fifth
approach we make the step size dependent on both the reduced mass
and the absolute value of the harmonic frequency associated with the
normal coordinate involved:
\begin{eqnarray}
q_{step}(i)=\hbox{constant}\times\sqrt{\frac{\mu}{\hbox{amu}}}\times \sqrt{\frac{1000~\hbox{cm}^{-1}}{\omega(i)}}\label{eq:method5}
\end{eqnarray}
Furthermore, in the absence of an analytical fourth derivatives code
which would by definition render the `true' answer, we can attempt
determining the latter by means of Richardson
extrapolation\cite{RalRab}. At the lowest level, we combine two
different step sizes h and 2h at the position x and reduce the error
like:
\begin{eqnarray}
D_1=\phi_{jjjj}(x)+O(h^2)=\frac{\omega_j(x+h)+\omega_j(x-h)-2\omega_j(x)}{(h)^2}\\
D_2=\phi_{jjjj}(x)+O(4h^2)=\frac{\omega_j(x+2h)+\omega_j(x-2h)-2\omega_j(x)}{(2h)^2}
\end{eqnarray}
and thus:
\begin{eqnarray}
\frac{(4D_1-D_2)}{3}=D_1+\frac{D_1-D_2}{3}=\phi_{jjjj}+O(h^4)
\end{eqnarray}
This formula is actually equivalent to the use of a five-point
fourth-order central difference formula. The Richardson approach can
be applied recursively in order to minimize discretization error:
with level $m$ estimates having an error term $O(h^{2m})$, the next
level Richardson estimate $D^{[m]}_1+(D^{[m]}_1-D^{[m]}_2)/(4^m-1)$ will
have an error $O(h^{2m+2})$.

We have tightened convergence criteria at all stages of the
electronic structure calculations to $10^{-10}$ or better (no
convergence could be achieved with even tighter criteria). Yet even
so, unacceptable roundoff error is introduced for small step sizes.
As a result, Richardson extrapolation is only reliable for rather
large step sizes, as can be seen in Table \ref{tab1}. The last column
shows the most accurate values, combining the step sizes of 0.10,
0.12 and 0.14 bohr. Even when including the step size of 0.08 bohr
(in column 8) into the formula, we probably see some roundoff error,
explaining the (small) difference of up to 3 cm$^{-1}$ between both
values --- nevertheless, we expect the algorithm to be nearly
converged.
Comparing methods (1-5, method 1 with a stepsize of 0.025 \AA ), the
maximum (mean absolute) errors compared to the last column
(our reference, in fact) are 4 (1.6), 4 (1.2), 9 (4.0), 8 (2.1)
and 1 (0.2) cm$^{-1}$. All methods (except method 3) have been calculated by
finding an optimal constant stepsize prefactor, which is shown in
the second line of Table \ref{tab1}.
This constant prefactor is very important which becomes apparent when
looking at mode 1 and mode 7 to 9 of method 1: For large frequencies
(like mode 1), a small step size is needed as can be seen when
comparing to the reference values. The error at 0.12 bohr for this
mode will be as large as 30 cm$^{-1}$. On the other hand, a step
size of 0.02 bohr yields errors of 50 cm$^{-1}$ for mode 7 and
30 cm$^{-1}$ for mode 8. The spurious interchange of fundamentals
8 and 9 at this step size could even lead to an incorrect assignment.
Thus, the choice of step size is critical when calculating quartic
force fields, and a worst-case error of about 5 cm$^{-1}$ still
needs to be assumed even with all step size algorithms.
These step size issues in fact transfer to $all$ molecules investigated,
as we have done also with method (4).
Here, the force fields are (compared to our reference
coupled cluster values), as a rule of thumb, about twice as bad as
when using method (5).
The only fundamental remedy would be the calculation of third
Kohn-Sham derivatives and taking numerical first derivatives of
these, which to our knowledge has not been implemented in any
quantum chemical code.

\subsection{Exchange-correlation functionals}

For assessing the different currently available density functionals
for this problem, we employed several GGA functionals
(BLYP\cite{BLYP}, HCTH407\cite{HCTH407}, PBE \cite{PBE}) and hybrid
density functionals (B3LYP\cite{B3LYP}, B97-1\cite{HCTH93},
B97-2\cite{B97-2}, and PBE0\cite{PBE0}). For all these functionals,
we consider the full validation set of 17 molecules mentioned above.

For this purpose, we used both the TZ2P and cc-pVTZ basis sets to
look at the performance of the functionals for the different basis
sets. Although force fields for all the aforementioned molecules
have been calculated with the two basis sets mentioned, the
C$_2$H$_2$ molecule required diffuse functions to yield
qualitatively correct bending anharmonicities. This is consistent
with results obtained by {\it ab initio} correlation methods (which
generate spurious positive anharmonicities with standard basis
sets), whereas Hartree-Fock renders negative anharmonic corrections
even without the use of diffuse functions\cite{c2h2}. In all tables
and figures, results for C$_2$H$_2$ thus refer to the aug-cc-pVTZ
basis set. In the case of the symmetric tops CH$_4$, SiH$_4$ and
SiF$_4$, we perturbed the masses of the hydrogens by 0.0001 atomic
mass units in order to break symmetry. Here, Coriolis coupling
between the now non-degenerate frequencies had to be taken into
account. For CH$_2$, a quasi-linear molecule, both B97-1 and
HCTH/407 yield positive anharmonic corrections when using the TZ2P
basis set- those were the only calculations excluded from our
evaluation. For CH$_2$NH, we were unable to use {\sc spectro} due to
the complicated resonances in this molecule (see Ref.\cite{ch2nh}
for a discussion)  --- the 
deperturbed values of the 9$\times$9
perturbation matrix from {\sc polyad} were consistently used instead.

\begin{figure}
\includegraphics[width=12cm,angle=270]{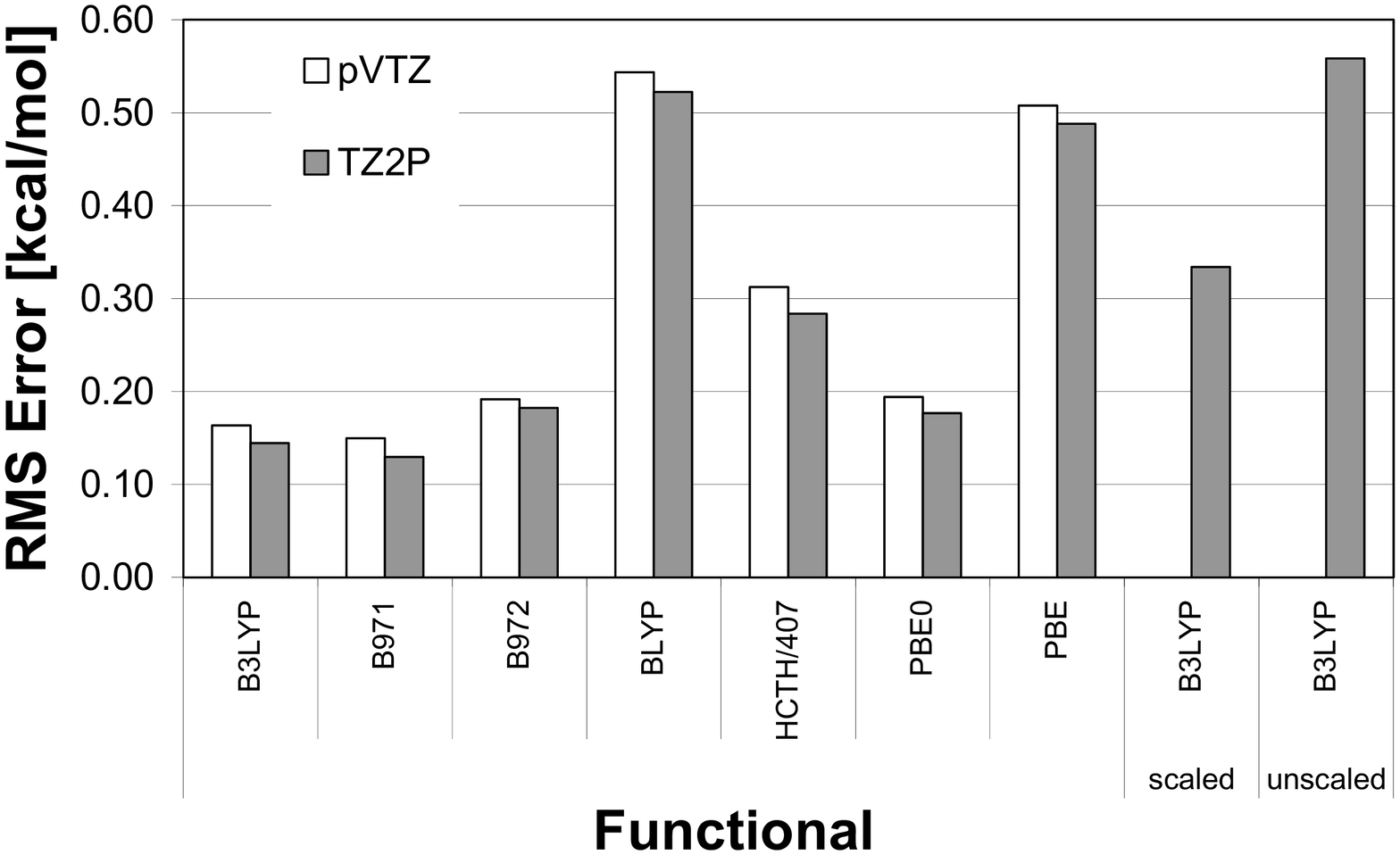} \\
\caption{\label{fig3}Boese et al}
\end{figure}

In figure \ref{fig3}, the RMS
and mean absolute errors of the zero-point energies for all the
molecules mentioned above are shown. Discussing the GGA functionals,
HCTH/407 is a vast improvement over both PBE and BLYP, reducing
their RMS error by almost a factor of two. Still, it is not quite
comparable to the hybrid functionals, which render the lowest
errors. The error in their zero-point energies is another factor of
two lower than the one obtained by HCTH/407. The scaled B3LYP
harmonic zero-point energy displays an error compared to HCTH/407.
The error is about 0.33 kcal/mol, compared to 0.13 to 0.19 kcal/mol
for the hybrid functionals. Here, B97-1 and B3LYP yield the lowest
errors around 0.14 kcal/mol for both basis sets. For the cc-pVTZ
basis set, all functionals yield a marginally worse zero-point
energy compared to the TZ2P basis set. This is in line with our
previous studies on geometries\cite{BMH}, where the combination of
DFT/cc-pVTZ generally showed larger errors compared than DFT/TZ2P.
Generally, the B3LYP functional returns results about 50\% better
than the ZPE determined by scaled B3LYP.

Discussing the fundamental frequencies in table \ref{tab2}, we can
evaluate the accuracy of density functionals for the calculation of
such properties which might prove useful when assigning experimental
IR-spectra. All hybrid functionals yield RMS errors between 30 and
40 cm$^{-1}$ for the 103 frequencies investigated, with B97-1/TZ2P
showing consistently the lowest errors. All pure GGA functionals
underestimate both harmonic and fundamental frequencies by at least
25 cm$^{-1}$, casting doubt on any values computed by such methods.
The difference between the errors of the calculated anharmonic and
harmonic frequencies, shown in the first four columns, which are
almost negligible and very large in comparison to those just done in
the correction (last two columns). The latter emphasizes the
agreement when using accurate harmonic frequencies from a higher
level method such as CCSD(T). For such a combined method, the RMS
errors are between 6 and 11 cm$^{-1}$, with surprisingly the hybrid
functionals such as B97-1 and PBE0 having errors around 9 cm$^{-1}$
and HCTH/407 and BLYP around 6 cm$^{-1}$. In comparison, determining
the anharmonic correction by simply scaling the harmonic frequency
by a constant factor fails completely: Its RMS error is as large as
for the fundamental frequencies itself.

As for the zero-point energies, HCTH/407 is slightly more accurate
than the scaled B3LYP harmonic frequencies, and BLYP is somewhat
less accurate than PBE for our validation set. Overall, we can
expect an accuracy of about 30 cm$^{-1}$ or less when calculating
fundamental frequencies for a diverse set of semirigid molecules
using perturbation theory. This is much more than we would expect
for organic molecules\cite{Azabenzenes}, but consistent with
previous studies on a larger number of harmonic
frequencies\cite{BMK}.

\begin{figure}
\includegraphics[width=12cm,angle=270]{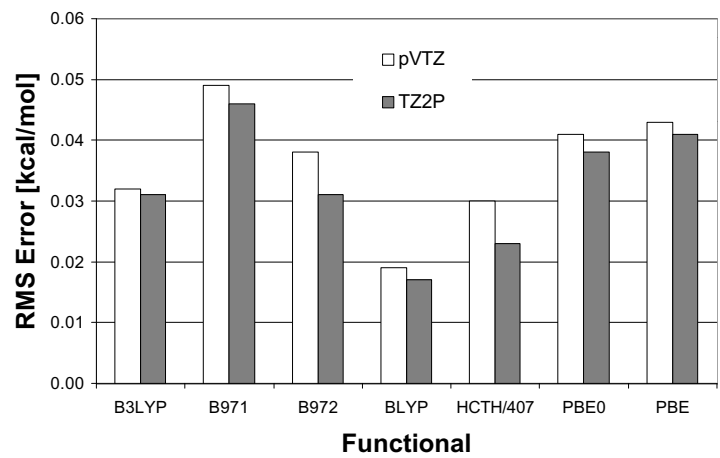} \\
\caption{\label{fig4}Boese et al}
\end{figure}

Returning to zero-point energies, these will be much improved when
including harmonic frequencies at the reference level and just
adding DFT anharmonic corrections. These errors are displayed in
Figure \ref{fig4}, and can be readily compared to those obtained in
Figure \ref{fig3} (note the different scale of both figures). The
error is reduced by at least a factor of five, showing the
superiority of such a combined method. As for the anharmonic
corrections, the GGA functionals BLYP and HCTH/407 yield the lowest
errors, about reducing the errors of the hybrid functionals by 50\%.
In comparison, simply using the scaled B3LYP frequencies shows no
improvement, as the RMS error would remain unchanged at 0.33
kcal/mol.

An additional observable obtained when calculating the anharmonic
force field is the rotational constant. While A$_e$, B$_e$ and C$_e$
are only dependent on the electronic minimum geometry, the
vibrationally averaged A$_0$, B$_0$ and C$_0$ are accessible by
experiment. In Table \ref{tab3}, the mean and RMS errors (in \%) for the 37
symmetry-unique rotational constants of the 16 molecules
investigated are displayed. 
For the correction $A_e-A_0$ we chose not to report the errors in \%, since
the reference values are already very small and thus large relative errors are 
made by the functionals.
As expected, the GGA functionals fare
worse than the hybrid functionals with the exception of the HCTH/407
functional. This exceptional property of HCTH/407 has been reported
before\cite{BMK,tHCTH}, with its geometry errors reduced by half in
comparison to the other GGA functionals. For the hybrid functionals,
the cc-pVTZ basis set gives lower errors than TZ2P, with the
opposite for all GGA functionals. The accuracy of the hybrid
functionals range from 0.06 to 0.13 cm$^{-1}$ 
(0.9 to 1.25 \%) for $A_e$ and $A_0$.
When only comparing
the correction R$_e$-R$_0$ to these values, the error will be cut in
half. This contrasts the RMS errors of the frequencies, where the
anharmonic correction has an error which is about five times as low
as the error of the frequency itself. Thus, adding the corrections
for example to CCSD(T) structures will not be as rewarding as it was
the case in the previous paragraph. As in Table \ref{tab2} for the
anharmonic corrections, all functionals now give quite similar
errors.

\subsection{Thermodynamic functions}

Finally, we turn to thermodynamic functions evaluated by four
different methods and compare them to our reference values. Since
rotational and vibrational levels contribute to those, error
cancellation can take effect here.

In Table \ref{tab4} (method 1), the heat capacity, the difference of
the enthalpy at a given temperature and the enthalpy at 0K, and the
entropy are given at room temperature, 600 K and 2000 K. Whereas
there is little difference between both basis sets, the functionals
show the expected behavior, with the hybrid functionals and HCTH/407
giving quite similar errors. BLYP and PBE are again the worst
performers, their errors being about twice as large. For the heat
capacity, the initial error at 298.15 K is larger than the error at
600 K for all functionals, suggesting some error cancellation at this
value. 

The relative importance of various post-RRHO contributions at high temperature 
(say, 2000 K) varies
somewhat with the molecule. For H$_2$O, vibrational anharmonicity and centrifugal
stretching are about equally important, rovibrational coupling
rather less so. For CH$_2$NH, vibrational anharmonicity predominates, with 
rovibrational coupling a distant second. For both C$_2$H$_2$ and N$_2$O,
vibrational anharmonicity far outweighs the two other contributions. In SO$_2$
a balance between the three contributions prevails.

When applying the RRHO approximation using just harmonic
frequencies and bottom-of-the-well rotational constants, as reported
in Table \ref{tab5} (method 2), somewhat surprising results are
obtained: At low temperatures (room temperature), this method is
easily competitive to calculating the full force field. In fact,
most errors will be lower when using the RRHO approximation in
conjunction with DFT for this temperature\cite{TSEliot}. Especially
the errors of BLYP and PBE are reduced and comparable to the hybrid
functionals for all temperatures. Again, there is barely any
difference between the two basis sets. At 600 K, the entropy and
enthalpy show the same behavior, only the heat capacity already has
an error twice as high than when calculating the force field. At
2000 K, the errors of the RRHO approximation becomes quite large,
with a factor of two (for the entropy) up to five (for enthalpy and
heat capacity) between Table \ref{tab4} and Table \ref{tab5}. Thus,
when calculating such properties at room temperature, the error in
the DFT method itself will be large compared to the error by making
such an approximation. Taking just the coupled cluster frequencies
and rotational constants from the reference method shows a similar
behavior, with the errors naturally reduced in comparison to DFT by
another factor of two for 298.15 K. At 600K, the error becomes
already comparable to the one obtained by the best density
functionals, and at 2000 K it is even higher in many cases. 

Table \ref{tab6} (method 3) represents a common compromise between
the RRHO approximation and using the full scale quartic force field
--- especially when working from experimental data. Here, anharmonic
frequencies and vibrationally averaged rotational constants are
substituted in the RRHO approximation and all additional effects of
anharmonicity, rovibrational coupling, and centrifugal stretching
are neglected. At low temperatures, this method basically yields the
same results as fully anharmonic calculations.
The difference between both methods is less than 5\% at room
temperature and 600 K, with BLYP and PBE yielding again larger
errors than the hybrid functionals. At 2000 K, however, neglecting
these terms causes a large loss in accuracy, although the errors are
not as large as when completely neglecting anharmonic effects.
Still, the error in the heat capacity (normally the most sensitive
quantity, as it involves the second moment of the partition
function) is only lowered by about 20\%, and the error of the
entropy and enthalpy approximately by a third. Method 3 is thus an
acceptable solution at low temperatures, and a stop-gap solution at
elevated temperatures (around 600 K) when no quartic force field is
available. By replacing the density functional values by the ones of
our reference method, the error becomes much smaller. Again, an
estimate can be made when this method becomes unreliable -- this
will be only at very large temperatures above 600 K. Of course, it
requires the calculation of a full CCSD(T) force field which is done
in our reference calculations, which would be prohibitive for many
molecules.

Finally, the last method (4), shown in Table \ref{tab7}, includes
all the effects covered in method 1, but in addition the equilibrium
geometry and harmonic frequencies from the reference coupled cluster
calculation were substituted in the rovibrational perturbation
theory analysis. (This represents a hybrid approach in which a
harmonic large basis set coupled cluster calculation would be
combined with a DFT anharmonic force field.)
Comparing the results to Table \ref{tab4}, especially at low
temperatures an improvement is apparent. The initial error is
reduced by a factor of two to three at room temperature. For larger
temperatures, this improvement becomes visibly smaller, with the
entropy the most affected variable by the change. At 2000 K, it is
the only function showing a lower error than the one obtained by a
pure DFT anharmonic force field. For both the heat capacity and the
enthalpy, an error cancellation for DFT takes place, thus showing a
lower error for method (1) than method (4). Comparing the
functionals, only PBE gives somewhat worse results for this method.
For the heat capacities and enthalpy at 298.15 and 600 K, the
accuracy of this method can already be compared to the anharmonic
CCSD(T) results obtained by method(3). The entropy, however will
yield errors about four times as large at 298.15 K. This is of
course the only method which yields a reasonable performance over
the full temperature range.

\section{Conclusions}

In this validation study, we have calculated DFT anharmonic force fields for
17 small (triatomic and larger) molecules.

At least using the present approach (finite differences of analytical Hessians), 
we find anharmonicities to be very sensitive to the DFT integration grid, with 
grids as large as 140$\times$974 (first row) and 200$\times$974 (second row)
being required for 1 cm$^{-1}$ numerical precision in fundamental frequencies.

The finite difference step sizes are another source of error. 
Either high-frequency or low-frequency modes are mainly affected, and too large or
too small values will lead to large errors.
Various adaptive-stepsize approaches yield satisfactory results.

Basis sets of at least TZ2P quality appear to be called for, although DZP quality 
may be an acceptable compromise for the very largest molecules.

HCTH/407 appears to be the most suitable GGA functional for the purpose, and
B97-1 the most suitable hybrid functional, immediately followed by B3LYP. Somewhat 
surprisingly, when combining DFT anharmonicities with large basis set CCSD(T) 
geometries and harmonic frequencies, GGA functionals yield better results
than their hybrid counterparts.

Fundamental frequencies can be determined to an accuracy of only 30
cm$^{-1}$, which is about the error that functionals also yield for
harmonic frequencies. Of course, for organic molecules, this error
will be reduced and DFT might prove useful in helping
experimentalists in their assignments. For inorganic molecules,
however, this accuracy might not be enough, and a higher level
method will have to be used to calculate the harmonic frequencies.

The same applies to zero-point energies. DFT zero-point energies
alone might not be accurate enough for using them as an addition to
W2 or W3 theories, only in combination with e.g. CCSD(T) harmonic
frequencies their values will get the desired error estimations.
Since a full CCSD(T) force field with a sufficiently large basis set
is so expensive, this might be
an attractive alternative in many cases.

DFT may be useful in computing vibrational corrections to rotational 
constants obtained at higher levels of theory: the intrinsic errors in 
DFT rotational constants are large enough that they outweigh any advantage
gained by the anharmonic computation.

DFT anharmonic force fields are useful in obtaining thermodynamic 
functions at elevated temperatures. At room temperature, the RRHO 
approximation is generally sufficient for semirigid molecules.

Finally, DFT-computed anharmonic corrections to the zero-point vibrational energy
will somewhat enhance the reliability of high-level ab initio thermochemical data.

\section{Acknowledgments}
ADB acknowledges a postdoctoral fellowship from the Feinberg
Graduate School (Weizmann Institute). Research at Weizmann was
supported by the Minerva Foundation, Munich, Germany, by the Lise
Meitner-Minerva Center for Computational Quantum Chemistry (of which
JMLM is a member), and by the Helen and Martin Kimmel Center for
Molecular Design. This work is related to Project 2003-024-1-100, "Selected Free Radicals and Critical Intermediates: Thermodynamic Properties from Theory and Experiment," of the International Union of Pure and Applied Chemistry (IUPAC).

 \indent

\newpage
\pagestyle{empty} \clearpage

\newpage
\pagestyle{empty} \clearpage

\newpage
\pagestyle{empty} \clearpage

\newpage
\pagestyle{empty} \clearpage

\clearpage \noindent

Fig. 1:\\
Error in the off-diagonal anharmonicity between both stretches. The
absolute values for these values at the reference grids are for 172
cm$^{-1}$ (H$_2$O), 12 cm$^{-1}$ (SO$_2$) and 24 cm$^{-1}$ (N$_2$O).

Fig. 2:\\
Basis set error, together with overall mean error, for the
C$_2$H$_4$, H$_2$O, SO$_2$ and N$_2$O molecules using different
basis sets. The white bars show basis set truncation error relative
to aug-cc-pVQZ, while the grey bars show the error compared to the
reference values (large basis set CCSD(T), see references).

Fig. 3:\\
RMS errors of several functionals for the anharmonic ZPVE, compared
to to scaled and unscaled harmonic B3LYP values.

Fig. 4:\\
RMS errors of several functionals for the anharmonic ZPVE, combining
CCSD(T) harmonic frequencies and DFT anharmonic corrections.

\newpage
\clearpage
\begin{table}
\caption{B97-1/TZ2P fundamental frequencies of CH$_2$NH using
different numerical differentiation step sizes. All step sizes are in multiples of
0.01 a.u. unless indicated otherwise.\label{tab1}}
\begin{tabular}{|c|c c c|c|c|c|c|c c|} \hline\hline
Method  &\multicolumn{3}{c|}{ Uniform (1)}& (2) & (3) & (4) & (5)
&\multicolumn{2}{c|}{Richardson extrap.}
\\\hline

Mode    &  2   & 0.025\AA&   12 & 0.025\AA &       &  6   &  4   & 8/10/12/14   &
10/12/14 \\\hline
   1    & 3279 & 3281 & 3252 & 3280  & 3275  & 3283 & 3283 &     3283     &     3284   \\
   2    & 2996 & 2996 & 2987 & 2995  & 2995  & 2995 & 2997 &     2997     &     2997   \\
   3    & 2868 & 2868 & 2860 & 2868  & 2868  & 2878 & 2869 &     2870     &     2870   \\
   4    & 1670 & 1672 & 1671 & 1672  & 1673  & 1672 & 1672 &     1672     &     1672   \\
   5    & 1458 & 1460 & 1461 & 1460  & 1462  & 1460 & 1460 &     1460     &     1460   \\
   6    & 1323 & 1335 & 1339 & 1336  & 1339  & 1338 & 1336 &     1335     &     1336   \\
   7    & 1091 & 1131 & 1140 & 1137  & 1141  & 1138 & 1135 &     1132     &     1135   \\
   8    & 1045 & 1073 & 1079 & 1077  & 1081  & 1079 & 1076 &     1074     &     1076   \\
   9    & 1058 & 1058 & 1060 & 1058  & 1064  & 1058 & 1058 &     1057     &     1058   \\
\hline\hline
\end{tabular}

\noindent (1) Uniform `one size fits all' step size.\\

\noindent (2) Step size proportional to reduced mass of normal mode, eq. (\ref{eq:method2})\\

\noindent (3) Step size chosen such that harmonic energy change will
amount to 1 millihartree for that mode.\\

\noindent (4) Using eq. (\ref{eq:method4}).\\

\noindent (5) Using eq. (\ref{eq:method5}).\\

\end{table}

\newpage
\clearpage
\begin{table}
\caption{Errors (cm$^{-1}$) in harmonic frequencies, fundamental
frequencies, and anharmonic corrections for several
exchange-correlation functionals and basis sets.\label{tab2}}
\begin{tabular}{|c c|c c|c c|c c|} \hline\hline
\multicolumn{2}{|c|}{ Property } &\multicolumn{2}{c|}{ Harmonic
Frequency }&\multicolumn{2}{c|}{ Fundamental Frequency }
&\multicolumn{2}{c|}{ Correction}
\\\hline

Method & Basis set & mean & RMS & mean & RMS & mean & RMS \\\hline
B3LYP   & TZ2P     &  -4  &  35 &  -2  &  35 & -2.4 &  7.7 \\
        & cc-pVTZ  &  -6  &  40 &  -4  &  39 & -1.5 &  6.1 \\
B97-1   & TZ2P     &  -6  &  32 &  -3  &  32 & -3.7 & 10.5 \\
        & cc-pVTZ  &  -7  &  37 &  -3  &  35 & -2.8 &  8.1 \\
B97-2   & TZ2P     &   8  &  33 &  11  &  38 & -2.4 &  6.7 \\
        & cc-pVTZ  &   8  &  36 &  11  &  39 & -2.9 &  7.8 \\
BLYP    & TZ2P     & -56  & 109 & -56  & 108 & -0.3 &  6.2 \\
        & cc-pVTZ  & -57  & 109 & -57  & 110 &  0.1 &  5.6 \\
HCTH/407& TZ2P     & -25  &  55 & -24  &  56 & -1.2 &  6.8 \\
        & cc-pVTZ  & -27  &  58 & -25  &  59 & -1.4 &  6.6 \\
PBE0    & TZ2P     &   7  &  35 &  11  &  40 & -3.6 &  9.3 \\
        & cc-pVTZ  &   7  &  40 &   9  &  43 & -2.9 &  8.2 \\
PBE     & TZ2P     & -51  &  93 & -49  &  92 & -2.3 &  9.3 \\
        & cc-pVTZ  & -52  &  93 & -49  &  93 & -1.6 &  7.0 \\
scaled B3LYP& cc-pVTZ &      &     &  24  &  61 & -30  & 68   \\
\hline\hline
\end{tabular}
\end{table}

\newpage
\clearpage
\begin{table}
\caption{Errors (cm$^{-1}$) for equilibrium and vibrational ground
state rotational constants (and difference between them) for several
exchange-correlation functionals and basis sets. For the TZ2P basis
set, the CH$_2$ molecule has been excluded in the
evaluation. 
In case of the total values of the rotational constants, all values
have been reported in \%, with the RMS error corresponding to the RMS
error of all the individual \% errors.
\label{tab3}}
\begin{tabular}{|c c|c c|c c|c c|} \hline\hline
\multicolumn{2}{|c|}{ Property } &\multicolumn{2}{c|}{ $B_e$ (\%) }
&\multicolumn{2}{c|}{ $B_0$ (\%) } &\multicolumn{2}{c|}{ Correction }
\\\hline

Method & Basis set & mean &  RMS & mean & RMS  & mean & RMS \\\hline
B3LYP   & TZ2P     &-0.18 & 1.05 & 0.14 & 1.12 & 0.015 & 0.047 \\
        & cc-pVTZ  &-0.31 & 1.01 &-0.16 & 1.10 & 0.008 & 0.037 \\
B97-1   & TZ2P     &-0.42 & 1.01 &-0.19 & 1.03 & 0.012 & 0.043 \\
        & cc-pVTZ  &-0.52 & 1.11 &-0.33 & 1.22 & 0.009 & 0.037 \\
B97-2   & TZ2P     & 0.34 & 0.92 & 0.61 & 1.15 & 0.015 & 0.047 \\
        & cc-pVTZ  & 0.34 & 0.94 & 0.54 & 1.13 & 0.010 & 0.038 \\
BLYP    & TZ2P     &-2.28 & 2.77 &-2.19 & 2.81 & 0.014 & 0.051 \\
        & cc-pVTZ  &-2.30 & 2.79 &-2.31 & 2.86 &-0.002 & 0.034 \\
HCTH/407& TZ2P     &-0.61 & 1.25 &-0.50 & 1.30 & 0.005 & 0.040 \\
        & cc-pVTZ  &-0.62 & 1.39 &-0.50 & 1.39 & 0.005 & 0.029 \\
PBE0    & TZ2P     & 0.25 & 0.90 & 0.59 & 1.19 & 0.018 & 0.051 \\
        & cc-pVTZ  & 0.25 & 0.95 & 0.39 & 1.11 & 0.010 & 0.041 \\
PBE     & TZ2P     &-2.01 & 2.32 &-1.84 & 2.28 & 0.018 & 0.057 \\
        & cc-pVTZ  &-2.02 & 2.42 &-1.87 & 2.40 & 0.005 & 0.028 \\
\hline\hline
\end{tabular}

\end{table}

\newpage
\clearpage
\begin{table}
\caption{RMS errors for thermodynamic functions at several
temperatures using DFT anharmonic force fields\label{tab4}}
\begin{tabular}{|c c|c c c|c c c|c c c|} \hline\hline
\multicolumn{2}{|c|}{ Property } &\multicolumn{3}{c|}{Heat capacity}
&\multicolumn{3}{c|}{Enthalpy function}
&\multicolumn{3}{c|}{Entropy}
\\\hline
\multicolumn{2}{|c|}{ } &\multicolumn{3}{c|}{ $C_p$ [J/K.mol] }
&\multicolumn{3}{c|}{ $H-H_0$ [kJ/mol] } &\multicolumn{3}{c|}{ $S$
[J/K.mol] }
\\\hline

Functional & Basis set & 298.15 & 600 & 2000 & 298.15 & 600 & 2000 &
298.15 & 600 & 2000 \\\hline
B3LYP   & TZ2P     & 0.57  & 0.42 & 0.92 & 0.08  & 0.23 & 0.85 & 0.53 &
0.83 & 1.23 \\
        & cc-pVTZ  & 0.61  & 0.40 & 0.67 & 0.09  & 0.23 & 0.64 & 0.95 &
1.10 & 1.33 \\
B97-1   & TZ2P     & 0.60  & 0.41 & 0.56 & 0.09  & 0.24 & 0.71 & 0.52 &
0.86 & 1.20 \\
        & cc-pVTZ  & 0.64  & 0.41 & 0.69 & 0.10  & 0.24 & 0.70 & 0.54 &
0.84 & 1.17 \\
B97-2   & TZ2P     & 0.47  & 0.45 & 0.67 & 0.06  & 0.24 & 0.78 & 0.44 &
0.82 & 1.25 \\
        & cc-pVTZ  & 0.54  & 0.46 & 0.78 & 0.08  & 0.25 & 0.81 & 0.46 &
0.87 & 1.30 \\
BLYP    & TZ2P     & 0.93  & 0.96 & 1.10 & 0.14  & 0.43 & 1.48 & 1.08 &
1.69 & 2.51 \\
        & cc-pVTZ  & 0.91  & 0.89 & 1.13 & 0.15  & 0.39 & 1.42 & 1.05 &
1.57 & 2.34 \\
HCTH/407& TZ2P     & 0.65  & 0.66 & 0.66 & 0.10  & 0.30 & 1.00 & 0.63 &
1.08 & 1.65 \\
        & cc-pVTZ  & 0.68  & 0.65 & 0.99 & 0.10  & 0.29 & 0.93 & 0.64 &
1.02 & 1.53 \\
PBE0    & TZ2P     & 0.47  & 0.44 & 1.02 & 0.07  & 0.20 & 1.00 & 0.44 &
0.71 & 1.30 \\
        & cc-pVTZ  & 0.55  & 0.45 & 0.76 & 0.08  & 0.22 & 0.78 & 0.47 &
0.76 & 1.21 \\
PBE     & TZ2P     & 0.90  & 0.93 & 1.03 & 0.14  & 0.41 & 1.28 & 0.94 &
1.52 & 2.22 \\
        & cc-pVTZ  & 0.94  & 0.91 & 1.12 & 0.14  & 0.40 & 1.27 & 0.96 &
1.47 & 2.15 \\
\hline\hline
\end{tabular}
\end{table}

\newpage
\clearpage
\begin{table}
\caption{RMS errors for thermodynamic functions at several
temperatures using DFT and the RRHO approximation with equilibrium
geometries and harmonic frequencies\label{tab5}}
\begin{tabular}{|c c|c c c|c c c|c c c|} \hline\hline
\multicolumn{2}{|c|}{ Property } &\multicolumn{3}{c|}{Heat capacity}
&\multicolumn{3}{c|}{Enthalpy function}
&\multicolumn{3}{c|}{Entropy}
\\\hline
\multicolumn{2}{|c|}{ } &\multicolumn{3}{c|}{ $C_p$ [J/K.mol] }
&\multicolumn{3}{c|}{ $H-H_0$ [kJ/mol] } &\multicolumn{3}{c|}{ $S$
[J/K.mol] }
\\\hline

Functional & Basis set & 298.15 & 600 & 2000 & 298.15 & 600 & 2000 &
298.15 & 600 & 2000 \\\hline
B3LYP   & TZ2P     & 0.53  & 0.78 & 2.84 & 0.07  & 0.25 & 2.58 & 0.46 & 0.85 & 2.45 \\
        & cc-pVTZ  & 0.54  & 0.77 & 2.82 & 0.07  & 0.25 & 2.55 & 0.88 & 1.14 & 2.57 \\
B97-1   & TZ2P     & 0.44  & 0.71 & 2.72 & 0.06  & 0.21 & 2.49 & 0.40 & 0.72 & 2.26 \\
        & cc-pVTZ  & 0.46  & 0.71 & 2.82 & 0.06  & 0.22 & 2.50 & 0.42 & 0.76 & 2.30 \\
B97-2   & TZ2P     & 0.56  & 0.83 & 2.91 & 0.07  & 0.24 & 2.72 & 0.45 & 0.81 & 2.60 \\
        & cc-pVTZ  & 0.60  & 0.84 & 2.89 & 0.08  & 0.26 & 2.70 & 0.49 & 0.88 & 2.62 \\
BLYP    & TZ2P     & 0.73  & 0.54 & 2.60 & 0.12  & 0.31 & 1.81 & 0.93 & 1.37 & 1.82 \\
        & cc-pVTZ  & 0.66  & 0.46 & 2.59 & 0.12  & 0.28 & 1.77 & 0.88 & 1.26 & 1.70 \\
HCTH/407& TZ2P     & 0.47  & 0.43 & 2.62 & 0.08  & 0.20 & 2.04 & 0.53 & 0.82 & 1.72 \\
        & cc-pVTZ  & 0.49  & 0.48 & 2.70 & 0.08  & 0.15 & 2.07 & 0.54 & 0.55 & 1.83 \\
PBE0    & TZ2P     & 0.59  & 0.84 & 2.90 & 0.08  & 0.31 & 2.71 & 0.48 & 1.13 & 2.64 \\
        & cc-pVTZ  & 0.63  & 0.89 & 2.88 & 0.08  & 0.37 & 2.68 & 0.52 & 1.43 & 2.67 \\
PBE     & TZ2P     & 0.61  & 0.44 & 2.61 & 0.11  & 0.26 & 1.76 & 0.79 & 1.13 & 1.59 \\
        & cc-pVTZ  & 0.60  & 0.46 & 2.58 & 0.11  & 0.26 & 1.72 & 0.78 & 1.15 & 1.55 \\
CCSD(T) &          & 0.28  & 0.75 & 3.01 & 0.03  & 0.18 & 2.74 & 0.26 & 0.57 & 2.50 \\
\hline\hline
\end{tabular}
\end{table}

\newpage
\clearpage
\begin{table}
\caption{RMS errors for thermodynamic functions at several
temperatures using DFT and the RRHO approximation with zero-point
average geometries and fundamental frequencies\label{tab6}}
\begin{tabular}{|c c|c c c|c c c|c c c|} \hline\hline
\multicolumn{2}{|c|}{ Property } &\multicolumn{3}{c|}{Heat capacity}
&\multicolumn{3}{c|}{Enthalpy function}
&\multicolumn{3}{c|}{Entropy}
\\\hline
\multicolumn{2}{|c|}{ } &\multicolumn{3}{c|}{ $C_p$ [J/K.mol] }
&\multicolumn{3}{c|}{ $H-H_0$ [kJ/mol] } &\multicolumn{3}{c|}{ $S$
[J/K.mol] }
\\\hline

Functional & Basis set & 298.15 & 600 & 2000 & 298.15 & 600 & 2000 &
298.15 & 600 & 2000 \\\hline
B3LYP   & TZ2P     & 0.57  & 0.44 & 2.44 & 0.08  & 0.22 & 1.78 & 0.53 & 0.85 & 1.60 \\
        & cc-pVTZ  & 0.60  & 0.45 & 2.45 & 0.09  & 0.23 & 1.80 & 0.95 & 1.14 & 1.80 \\
B97-1   & TZ2P     & 0.59  & 0.37 & 2.35 & 0.09  & 0.22 & 1.66 & 0.52 & 0.72 & 1.53 \\
        & cc-pVTZ  & 0.63  & 0.45 & 2.45 & 0.10  & 0.23 & 1.79 & 0.54 & 0.76 & 1.65 \\
B97-2   & TZ2P     & 0.46  & 0.50 & 2.50 & 0.06  & 0.23 & 1.94 & 0.43 & 0.81 & 1.82 \\
        & cc-pVTZ  & 0.54  & 0.54 & 2.52 & 0.08  & 0.25 & 1.98 & 0.46 & 0.88 & 1.93 \\
BLYP    & TZ2P     & 0.90  & 0.80 & 2.19 & 0.14  & 0.39 & 1.16 & 1.06 & 1.37 & 1.57 \\
        & cc-pVTZ  & 0.89  & 0.76 & 2.22 & 0.14  & 0.37 & 1.21 & 1.04 & 1.26 & 1.66 \\
HCTH/407& TZ2P     & 0.62  & 0.48 & 2.22 & 0.09  & 0.26 & 1.19 & 0.62 & 0.82 & 1.12 \\
        & cc-pVTZ  & 0.67  & 0.59 & 2.33 & 0.10  & 0.28 & 1.42 & 0.63 & 0.55 & 1.35 \\
PBE0    & TZ2P     & 0.46  & 0.49 & 2.50 & 0.07  & 0.20 & 1.92 & 0.43 & 1.13 & 1.75 \\
        & cc-pVTZ  & 0.55  & 0.53 & 2.51 & 0.08  & 0.22 & 1.93 & 0.47 & 1.43 & 1.82 \\
PBE     & TZ2P     & 0.49  & 0.82 & 2.19 & 0.06  & 0.39 & 1.21 & 0.93 & 1.13 & 1.57 \\
        & cc-pVTZ  & 0.93  & 0.82 & 2.21 & 0.14  & 0.38 & 1.26 & 0.95 & 1.15 & 1.66 \\
CCSD(T) &          & 0.07  & 0.29 & 2.49 & 0.00  & 0.06 & 1.78 & 0.03 & 0.14 & 1.33 \\
 \hline\hline
\end{tabular}
\end{table}

\newpage
\clearpage
\begin{table}
\caption{RMS errors for thermodynamic functions at several
temperatures using DFT anharmonic force fields combined with large
basis set CCSD(T) geometries and harmonic frequencies\label{tab7}}
\begin{tabular}{|c c|c c c|c c c|c c c|} \hline\hline
\multicolumn{2}{|c|}{ Property } &\multicolumn{3}{c|}{Heat capacity}
&\multicolumn{3}{c|}{Enthalpy function}
&\multicolumn{3}{c|}{Entropy}
\\\hline
\multicolumn{2}{|c|}{ } &\multicolumn{3}{c|}{ $C_p$ [J/K.mol] }
&\multicolumn{3}{c|}{ $H-H_0$ [kJ/mol] } &\multicolumn{3}{c|}{ $S$
[J/K.mol] }
\\\hline

Functional & Basis set & 298.15 & 600 & 2000 & 298.15 & 600 & 2000 &
298.15 & 600 & 2000 \\\hline
B3LYP   & TZ2P     & 0.17  & 0.29 & 0.90 & 0.03  & 0.08 & 1.05 & 0.18 & 0.25 & 0.90 \\
        & cc-pVTZ  & 0.18  & 0.26 & 0.81 & 0.03  & 0.08 & 0.94 & 0.19 & 0.24 & 0.81 \\
B97-1   & TZ2P     & 0.30  & 0.30 & 0.93 & 0.05  & 0.12 & 0.99 & 0.25 & 0.40 & 0.93 \\
        & cc-pVTZ  & 0.31  & 0.29 & 0.93 & 0.05  & 0.12 & 0.98 & 0.27 & 0.39 & 0.93 \\
B97-2   & TZ2P     & 0.09  & 0.30 & 0.93 & 0.01  & 0.12 & 0.99 & 0.14 & 0.39 & 0.93 \\
        & cc-pVTZ  & 0.09  & 0.30 & 0.96 & 0.01  & 0.12 & 1.01 & 0.14 & 0.39 & 0.96 \\
BLYP    & TZ2P     & 0.11  & 0.31 & 0.87 & 0.01  & 0.07 & 1.02 & 0.16 & 0.22 & 0.87 \\
        & cc-pVTZ  & 0.10  & 0.28 & 0.86 & 0.01  & 0.07 & 1.04 & 0.13 & 0.16 & 0.86 \\
HCTH/407& TZ2P     & 0.18  & 0.31 & 0.99 & 0.03  & 0.09 & 1.12 & 0.16 & 0.27 & 0.99 \\
        & cc-pVTZ  & 0.19  & 0.30 & 0.93 & 0.03  & 0.09 & 1.06 & 0.19 & 0.26 & 0.93 \\
PBE0    & TZ2P     & 0.10  & 0.27 & 0.81 & 0.01  & 0.06 & 0.97 & 0.15 & 0.18 & 0.81 \\
        & cc-pVTZ  & 0.10  & 0.24 & 0.74 & 0.01  & 0.06 & 0.89 & 0.16 & 0.16 & 0.74 \\
PBE     & TZ2P     & 0.47  & 0.36 & 1.08 & 0.07  & 0.18 & 1.02 & 0.35 & 0.59 & 1.08 \\
        & cc-pVTZ  & 0.50  & 0.35 & 1.05 & 0.07  & 0.18 & 0.98 & 0.37 & 0.59 & 1.05 \\
\hline\hline
\end{tabular}
\end{table}

\end{document}